\documentstyle[preprint,aps]{revtex}
\begin{document}
\draft
\newcommand{\comm}[1]{\underline{\tt #1}}

\tightenlines
\newcommand{\atn}{\tan^{-1}}
\newcommand{\ch}{{\rm ch}}
\newcommand{\sh}{{\rm sh}}
\newcommand{\th}{{\rm th}}
\newcommand{\bml}{\begin{mathletters}}
\newcommand{\eml}{\end{mathletters}}
\newcommand{\be}{\begin{equation}}
\newcommand{\ee}{\end{equation}}
\newcommand{\ba}{\begin{array}}
\newcommand{\ea}{\end{array}}
\newcommand{\bea}{\begin{eqnarray}}
\newcommand{\eea}{\end{eqnarray}}
\newcommand{\nn}{\nonumber}
\newcommand{\smi}{\!-\!}
\newcommand{\spl}{\!+\!}

\def\eqalign#1{
\null \,\vcenter {\openup \jot \ialign {\strut \hfil $\displaystyle {
##}$&$\displaystyle {{}##}$\hfil \crcr #1\crcr }}\,}

\pagestyle{myheadings}
\title{
{\flushright{\small cond-mat/9812415\\} \vspace{.2in}}
On the symmetry of excitations in SU(2) \\  
Bethe Ansatz systems}
\author{\large F.~Woynarovich}
\address{Institute for Solid State Physics and Optics\\
Hungarian Academy of Sciences\\
1525 Budapest 114, Pf 49.}
\maketitle
\begin{abstract}
Using the XXX Heisenberg chain as an example, based on the symmetry 
properties of the eigenstates with respect to reversing all the spins 
we argue, that the basic SU(2) symmetry of the model is inherited by 
the excitations with slight modifications only. 
\end{abstract}

\pacs{PACS numbers: 75.10.D}

\setlength{\parskip}{2ex}
\setlength{\parindent}{0em}
\setlength{\baselineskip}{3ex}

Recently several attempts have been made to define regularized 
field theoretical models as some limits of certain 
completly integrable lattice models \cite{Me,WoFo1,WoFo2,HaRaWo}.
In these works the limiting models are identified through the structure
and symmetry properties of the spectrum and the eigenstates. In some 
of the cases an SU(2) symmetry is present already in the initial lattice 
model \cite{WoFo1,WoFo2}, but in some this symmetry is recognized
through degenerations developing in the limiting process only
\cite{WoFo2,HaRaWo}. For this latter cases it can be instructive to see
models, in which the connection between the basic SU(2) symmetry of the 
model and the symmetry of the excitations can be dirrectly detected.
Our aim in the present note is to see this connection in the case of the 
isotropic Heisenberg chain, which is the symplest integrable SU(2) model. 

The XXX Heisenberg chain is given by the Hamiltonian
\be
\label{Ham} 
H=\sum_{j=1}^N \vec S_j\vec S_{j+1}=\sum_{j=1}^N
\left(S_j^xS_{j+1}^x+S_j^yS_{j+1}^y+
S_j^zS_{j+1}^z\right)\,. 
\ee 
The Hilbert space is the tensor product of $N$ spaces furnishing the
doublet representation of SU(2), $S_j^{x,y,z}$ are the spin operators
acting on the $j$th site, the $(N\!\!+\!\!1)$th site
is identified with the first one, and we suppose $N={\rm even}$. 
The Hamiltonian commutes with all the components of the total spin
\be
S^{x,y,z}=\sum_j S_j^{x,y,z}\,,
\ee
thus the eigenstates can be labelled by the values of the $S^{z}$ and $S^2$,
and the energy should show degenerations corresponding to SU(2) multiplets.
Another operator commuting with the Hamiltonian and being important in our
reasoning is 
\be\label{tukor}
\hat\Sigma=\prod_{j=1}^N\,\sigma_j^x\,, 
\ee 
which represents a reflection on the $x$-axis, but
in a basis given by the products of the $S^z$ eigenstates of the individual 
spins $\hat\Sigma$ symply flips all the spins (the up ones down and 
the down ones up). The $S^z=0$ 
eigenstates of (\ref{Ham}) are expected to be eigenstates of this operation: 
otherwise certain points of the spectrum were twofold degenerated. This 
actually can happen accidentally, but is not forced by the symmetry, as   
$\hat\Sigma$ has one-dimensional representations only. In case of 
normal SU(2) the eigenvalue of $\hat\Sigma$ can be given as 
\be\label{sigmaert}
\Sigma=(-1)^{N/2-l}
\ee
with $l$ defined by the spin-length $S^2=l(l+1)$. This is a consequence
of the fact, that any $S^2=l(l+1)$, $S^z=0$ state can be given as linear
combination of states built up as products of $l$ triplet and $N/2-l$
singlet pairs \cite{SuFa}.

The Hamiltonian (\ref{Ham}) has been diagonalized by Bethe Ansatz (BA)
\cite{Be,Or,desClGa,Yang}, and the BA equations (BAE) were analyzed
\cite{Wo1,BVV,ViWo,Wo2}.  
Due to these studies  the structure of the low energy states and the
spectrum are well known by now. The ground state is an SU(2) singlet, 
and as there is no parameter in it, it can be considered as a vacuum.
The excited states can be described as scattering states of dressed 
particles (spinwaves).
Each particle has a parameter usually called rapidity through which 
its energy
and momentum can be given. The total energy and momentum of a 
state are given by the sums of the contributions of the vacuum and the 
individual particles. The momenta of the $n$ particles are quantized 
through a set of $n+r$ equations of the BA type 
(higher level Bethe Ansatz equations). These equations determine in addition
to the 
particle momenta a set of $r(\leq n/2)$ variables, which do not enter either 
into the energy, or the momentum, but are needed to give the spin of the 
state:
\be\label{spin}
S^z=l\,,\quad {\rm and}\quad S^2=l(l+1)\quad {\rm with}\quad l={n\over2}-r\,,
\ee
These formulae and the fact, that the number of particles $n$ and the number 
of sites in the chain (the number of real spins) $N$ must be of the same 
parity suggest, that the spin of the 
individual particles is of length 1/2 \cite{FaTa}.

It is remarkable, that the BA solutions give the $S^2\!=\!S^z(S^z\!+\!1)$
((\ref{spin})), and the $S^z<l$ states should be constructed by the 
(repeted) application of the $\sigma^-$
operator ($\sigma^{\pm}=S^x\pm iS^y$). The states form normal SU(2)
multiplets (of the $N$ spins forming thye chain), and it was very tempting 
to consider them as normal SU(2)
multiplets of the dressed particles. This concept, however leads
to a contradiction with the eigenvalue of $\hat\Sigma$:
as the elements of a multiplet are $S^2=l(l+1)$ eigenstates of the $N$
spins forming the chain, the $S^z=0$ member is symmetric or antisymmetric 
under $\hat\Sigma$ according to the eigenvalue (\ref{sigmaert}), while
in case of particles obeying normal SU(2), this eigenvalue should be
$\Sigma=(-1)^{n/2-l}$, i.e.\ instead of the chainlength, the number of 
dressed particles should appear. Actually we think that $\Sigma$ 
factorizes as
\be\label{factor}
\Sigma=\Sigma_{vac.}\Sigma_{part.} 
\ee
with $\Sigma_{vac.}$ and $\Sigma_{part.}$ being the eigenvalues corresponding
to the vacuum and to the symmetry of the spin configuration of the particles, 
respectively:
\be\label{nsigmaert}
\Sigma_{vac.}=(-1)^{N/2}\,,\quad {\rm and}\quad\Sigma_{part.}=(-1)^{l}\,.  
\ee
(We think this is the case as if $\Sigma_{part.}$ can be defined, it should 
be independent of the chainlength.) We should emphasize, there is no 
contradiction between the spin-length and the spinreversal symmetry,
as long as we think in terms of $N$ spins, the contradiction arise if we 
want to interpret them as the spinlength and
the symmetry of the $n$ spins caried by the dressed particles.

Now we show, that the above contradiction can be dissolved supposing 
the particles obey a modified SU(2) equivalent to a $q$-deformed SU(2) at 
$q\!=\!-1$. Actually we show the structure in which 
the $S^z=0$ member of a $2l+1$-fold
degenerated multiplet is the eigenfunction of spin-reversal with the 
eigenvalue $\Sigma_{part.}$ of (\ref{nsigmaert}). 

Let us define the $\sigma$ operators of the $n$ ($n\!=\!{\rm even}$) spins
as
\be\label{szigmak}
\sigma^z=\sum_{j=1}^n\sigma_j^z\,,\quad 
\sigma^{+}=p\sum_{j=1}^n(-1)^{j-1}\sigma_j^{+}\,,\quad
\sigma^{-}=\sum_{j=1}^n(-1)^{j-1}\sigma_j^{-}\,,
\ee
with $p$ being $\pm1$ and  $\sigma_j$ acting on the spin of the $j$th particle. 
In the case $p=1$ these operators obey the normal SU(2) commutation 
relations, but correspond to an unusual comultiplication:
\be\label{komm}
\left\lbrack\sigma^+,\sigma^-\right\rbrack=\sigma^z\,,
\quad\quad
\left\lbrack\sigma^z,\sigma^{\pm}\right\rbrack=\pm2\sigma^{\pm}\,,
\ee
\be\label{comul}
\Delta(\sigma^z)=\sigma^z\otimes1+1\otimes\sigma^z\,,\quad\quad
\Delta(\sigma^{\pm})=\sigma^{\pm}\otimes1+(-1)^{\sigma^z}\otimes\sigma^{\pm}\,.
\ee
For $p=-1$ (\ref{szigmak}) correspond to a $q$-deformed SU(2) at 
$q\!=\!-1$ where the commutation relations an the cooproducts are
\be\label{kommq}
\left\lbrack\sigma^+,\sigma^-\right\rbrack=(-1)^{\sigma^z-1}\sigma^z\,,
\quad\quad
\left\lbrack\sigma^z,\sigma^{\pm}\right\rbrack=\pm2\sigma^{\pm}\,,
\ee
\bea\label{comulq}
\Delta(\sigma^z)&=&\sigma^z\otimes1+1\otimes\sigma^z\,,\nonumber\\
\Delta(\sigma^{+})&=&\sigma^{+}\otimes(-1)^{\sigma^z}+1\otimes\sigma^{+}\,,
\nonumber\\
\Delta(\sigma^{-})&=&\sigma^{-}\otimes1+(-1)^{\sigma^z}\otimes\sigma^{-}\,.
\eea
The two choices are equivalent, as redefining $\sigma^+$ as 
\be
\sigma^+\rightarrow(-1)^{\sigma^z-1}\sigma^+
\ee
interchanges the $p\!=\!1$ and $p\!=\!-\!1$ cases ($n\!=\!{\rm even}$). 
In the folloving we take
$p\!=\!1$.
The (\ref{szigmak}) operators 
can be obtained from the `normal' SU(2) $\sigma$s 
\be
\sigma^z_{SU(2)}=\sum_{j=1}^n\sigma_j^z\,,\quad 
\sigma^{\pm}_{SU(2)}=\sum_{j=1}^n\sigma_j^{\pm}\,.
\ee
by rotating evry second spin by $\pi$ around the $z$ axis, i.e.:
\be
\sigma^{z}=\sigma^{z}_{SU(2)}\,,\quad
\sigma^{\pm}=\hat O\hat\sigma^{\pm}_{SU(2)}\hat O^{-1}\,,
\ee
with
\be
\hat O=\exp\left\{i\pi\sum(j-1)\sigma_j^z/2\right\}\,.
\ee
As a consequence, states in the two Hilbert spaces can be connected so, that
the corresponding states have the same eigenvalues for 
$S^z$ and $S^2$ defined in their own
Hilbert space
($S^z=\sigma^z/2$, $S^2=\sigma^-\sigma^++S^z(S^z+1)$ resp.
$S^z=\sigma_{SU(2)}^z/2$, 
$S^2=\sigma_{SU(2)}^-\sigma_{SU(2)}^++S_{SU(2)}^z(S_{SU(2)}^z+1)$):
\be 
\vert\phi\rangle=\hat O\vert\phi\rangle_{SU(2)}\,.
\ee
As, however,
\be
\hat O^{-1}\hat\Sigma\hat O=(-1)^{n/2}\hat \Sigma\,,\quad\quad\quad
\left({\rm now}\quad\hat\Sigma=\prod_{j=1}^n \sigma_j^x\right)
\ee
the $S^z=0$ states have diferent eigenvalues under the spin reversal 
$\hat\Sigma$ (one corresponding to (\ref{nsigmaert}), the other 
being analogous to
(\ref{sigmaert})):
\be 
\hat\Sigma\vert\phi\rangle
=(-1)^l\vert\phi\rangle\quad{\rm and}\quad
\hat\Sigma\vert\phi\rangle_{SU(2)}
=(-1)^{n/2-l}\vert\phi\rangle_{SU(2)}\,.
\ee
(In symple terms: among the two particle states in the (\ref{szigmak})
structure contrary to the case of normal SU(2) 
the state $\frac{1}{\sqrt2}(|\uparrow\downarrow\rangle-
|\downarrow\uparrow\rangle)$ is the $S^z=0$ member of the triplet, and
the singlet is
$\frac{1}{\sqrt2}(|\uparrow\downarrow\rangle+
|\downarrow\uparrow\rangle)$.)

As a summary, based on the above we may say,
that even though we can not construct the spin operators of
the dressed particles, 
having $(2l+1)$-fold degenerated
multiplets where within a multiplet the states are distinguished by
a quantum number $S^z$ ($l\geq S^z={\rm integer}\geq-l$), the eigenvalue
$\Sigma=(-1)^{l}$ indicates that the spin space of the particles 
is of the above structure.

Finally we note, that we expect to be general, that in BA systens, where
the excited states can be described in terms of dressed particles, the 
underlying SU(2) symmetry of the system appears modified 
in the excitations: the contradiction to be 
dissolved between the $\Sigma$ for
$N$ particle building up the system and the $\Sigma$ for $n$ dressed
particles (namely that the first do not depend on $n$) is manifest in
other systems too regardless of the actual form of the Hamiltonian.

{\em Acknowledgements:}
I am grateful to J.~Balog, P.~Forg\'acs 
P.~Vecserny\'es and K.\ Szlach\'anyi for the illuminating discussions. 
The support from Hungarian National Science Fund OTKA under grant 
Nr.~T022607 is acknowledged.

\end{document}